\documentclass[aps,pr,superscriptaddress,preprintnumbers,nofootinbib,10pt]{revtex4-2}

\usepackage{multirow}
\usepackage{amsmath}
\usepackage{amssymb}
\usepackage[dvipdf,dvips]{graphicx}
\usepackage{color}
\usepackage{hyperref}
\usepackage{url}
\usepackage{slashed}
\usepackage{subfigure}
\usepackage[usenames,dvipsnames]{xcolor}
\usepackage{amsmath}
\usepackage{amsfonts}
\usepackage{float} 
\usepackage{amssymb}
\usepackage{epsfig}
\usepackage{graphics}
\usepackage{euscript}
\usepackage{slashed}
\usepackage{epstopdf}
\usepackage[utf8]{inputenc}
\allowdisplaybreaks
\usepackage[normalem]{ulem}
\usepackage{pifont}
\usepackage{dsfont}
\usepackage{MnSymbol}
\usepackage{verbatim}
\usepackage{graphicx}
\usepackage{latexsym}
\usepackage{tikz-feynman}
\usepackage{bbold}
\newcommand{\cA}{\cal A}
\newcommand{\cB}{\cal B}

\begin{document}

	\title{{\bf Bell-CHSH inequality and unitary operators}}

%	\author{P. De Fabritiis} \email{pdf321@cbpf.br} \affiliation{CBPF $-$ Centro Brasileiro de Pesquisas Físicas, Rua Dr. Xavier %Sigaud 150, 22290-180, Rio de Janeiro, Brazil}
	
%	\author{F. M. Guedes} \email{fmqguedes@gmail.com} \affiliation{UERJ $–$ Universidade do Estado do Rio de Janeiro,	Instituto de Física $–$ Departamento de Física Teórica $–$ Rua São Francisco Xavier 524, 20550-013, Maracanã, Rio de Janeiro, Brazil}
	
	\author{M. S.  Guimaraes}\email{msguimaraes@uerj.br} \affiliation{UERJ $–$ Universidade do Estado do Rio de Janeiro,	Instituto de Física $–$ Departamento de Física Teórica $–$ Rua São Francisco Xavier 524, 20550-013, Maracanã, Rio de Janeiro, Brazil}
	
	\author{I. Roditi} \email{roditi@cbpf.br} \affiliation{CBPF $-$ Centro Brasileiro de Pesquisas Físicas, Rua Dr. Xavier Sigaud 150, 22290-180, Rio de Janeiro, Brazil}
	
	\author{S. P. Sorella} \email{silvio.sorella@gmail.com} \affiliation{UERJ $–$ Universidade do Estado do Rio de Janeiro,	Instituto de Física $–$ Departamento de Física Teórica $–$ Rua São Francisco Xavier 524, 20550-013, Maracanã, Rio de Janeiro, Brazil}

	\begin{abstract}

Unitary operators are employed to investigate the violation of the Bell-CHSH inequality. The ensuing modifications affecting both classical and quantum bounds are elucidated. The relevance of a particular class of unitary operators whose expectation values are real is pointed out. For these operators, the classical and quantum bounds remain unaltered, being given, respectively, by $2$ and $2\sqrt{2}$.  As examples, we discuss the explicit realization of  phase space Bell-CHSH inequality violation and the Weyl unitary operators for a real scalar field in relativistic Quantum Field Theory. 

	\end{abstract}

	\maketitle

\section{Introduction}	

The Bell-CHSH inequality \cite{Bell64,CHSH69} is a fundamental tool in the study of the entanglement, a hallmark of Quantum Physics. In its usual form, the Bell-CHSH correlator reads 
\begin{equation} 
\langle \psi | \;{\cal C} | \;\psi \rangle \;, \label{corr}
\end{equation}
where $|\psi\rangle$ stands for the state of the system and ${\cal C}$ is given by 
\begin{equation} 
{\cal C} = (A + A') B + (A-A')B' \;, \label{cop}
\end{equation}
with $(A,A')$ and $(B,B')$ being the so-called Bell's operators \cite{Bell64}. They are dichotomic Hermitian operators fulfilling the following properties 
\begin{eqnarray} 
A & =& A^{\dagger} \;, \quad A'  = {A'}^{\dagger} \;, \quad B  = B^{\dagger} \;, \quad B'  = {B'}^{\dagger} \;, \nonumber \\[3mm]
A^2 & =& {A'}^2 = B^2 = {B'}^2=1 \;, \label{AB}
\end{eqnarray}
and
\begin{eqnarray}
\left[\; A,B\; \right] &= & 0\;, \quad \left[\; A',B\; \right] =  0 \;, \quad \left[\; A,B'\; \right] =  0 \;, \quad \left[\; A',B'\; \right] =  0 \;.  \nonumber \\[3mm]
\left[\; A,A'\; \right] &\neq & 0\;, \quad \left[\; B,B'\; \right] \neq  0 \;.   
\label{ABBell}
\end{eqnarray}
It is customary to refer the capital letters $A$ and $B$ as Alice and Bob. From equation \eqref{ABBell} one sees that Alice's operator commute with Bob's operators, while $(A,A')$ have nonvanishing commutator.  The same holds for $(B,B')$. This last condition has a simple understanding in terms of measurements. For instance, in the case of the usual spin $1/2$ Bell-CHSH inequality, it means that Alice and Bob cannot perform two simultaneous measurements of their own spin in two different directions.\\\\One speaks of a violation of the Bell-CHSH inequality whenever 
\begin{equation} 
2 < \big|  \langle \psi | \;{\cal C} | \;\psi \rangle \big| \le 2 \sqrt{2}  \;. \label{viol}
\end{equation}
The value $2$ is the classical bound while $2\sqrt{2}$, known as Tsirelson's bound \cite{TSI}, is the quantum upper bound. In order to better understand that bound, let us consider now $(A,A')$ and $(B,B')$ as generic operators and compute
\begin{eqnarray}\label{opBell}
	{\cal C}^{\dagger} {\cal C} = |{\cal C} |^2 =|A+A'|^2 |B|^2 +|A-A'|^2 |B'|^2 + (A^{\dagger}A - A'^{\dagger} A')(B^{\dagger}B' + B'^{\dagger} B) + (A'^{\dagger}A - A^{\dagger} A')(B^{\dagger}B' - B'^{\dagger} B)
\end{eqnarray}
where $|\cdots|$ denotes the non negative part of the polar decomposition of an operator.\footnote{The polar decomposition of an operator $A$ is $A = U|A|$, where $U$ is a partial isometry and $|A| = A^{\dagger} A$. If $A$ is a normal operator, $U$ is unitary and commutes with $|A|$.} The overall scales of the operators are not important to quantify a Bell test, that is why one uses dichotomic operators. Recalling that any operator can be cast in a polar decomposition, without losing in generality we demand that the non negative part of all operators are set equal to $1$, that is, $|A| = |A'| = |B| = |B'| = 1$. Therefore, expression \eqref{opBell} reduces to
\begin{eqnarray}
	|{\cal C}|^2 = 4 + (A'^{\dagger}A - A^{\dagger} A')(B^{\dagger}B' - B'^{\dagger} B)
\end{eqnarray}
the usual Tsirelson's bound is obtained  when the operators are also required to be Hermitian, so that the last term on the right hand side becomes a product of commutators $\left[ A', A\right] \left[ B, B' \right]$. In such a setting, each commutator has a maximum value of $2$, and in Quantum Mechanics a  state can be found that realizes this maximum value for both commutators, resulting in  $\langle |{\cal C}|^2  \rangle = 8$. The classical counterpart can be defined as the case where all variables commute and therefore the maximum value in this case is $\langle |{\cal C}|^2  \rangle = 4$. \\\\However,  we are interested here in the situation where operators belonging to the class of normal operators are employed in the  Bell-CHSH inequality. Normal operators have a spectral representation and they have been considered in ongoing discussions on the possibility of enlarging the class of the observables of the Quantum Theory \cite{Hu_2017,Erhard_2020}. An operator  ${\cal D} $ is called normal if it commutes with its hermitian conjugate, {\it i.e.} ${\cal D}^{\dagger} {\cal D} = {\cal D}{\cal D}^{\dagger}$. An Hermitian operator is a particular case of a normal operator. Note also that if the normal operator $ {\cal D} $ satisfies $| {\cal D} | = 1$ then it is a unitary operator. It follows that the relation \eqref{opBell} is valid for unitary operators, and these are therefore the natural class of normal operators to be used in Bell tests. \\\\The present paper aims at studying the Bell-CHSH inequality when unitary operators are employed instead of Bell's operators, namely, when 
\begin{equation} 
	{\cal C} = ({\cA}  + {\cA}') {\cB} + ({\cA} -{\cA}'){ \cB}' \;, \label{copun}
\end{equation}
where
\begin{eqnarray} 
	{\cA}^{\dagger} \cA & =  & 1 = \cA {\cA}^{\dagger}   \;, \quad   {{\cA}'}^{\dagger} {{\cA}'} =  1 = {\cA}' {{\cA}'}^{\dagger}  \;, \nonumber \\[3mm]
	{\cB}^{\dagger} \cB & =  & 1 = \cB {\cB}^{\dagger}   \;, \quad   {{\cB}'}^{\dagger} {{\cB}'} =  1 = {\cB}' {{\cB}'}^{\dagger} \;, \nonumber \\[3mm]
	\left[\; \cA, \cB\; \right] &= & 0\;, \quad \left[\; \cA', \cB\; \right] =  0 \;, \quad \left[\; \cA, \cB'\; \right] =  0 \;, \quad \left[\; \cA', \cB'\; \right] =  0 \;, \nonumber \\[3mm]
	\left[\; \cA, \cA'\; \right] &\neq & 0\;, \quad \left[\; \cB, \cB'\; \right] \neq  0 \;.
	\label{ABBellu}
\end{eqnarray}
Besides being the natural class of normal operators to be studied in Bell tests, one notices that in the realm of relativistic Quantum Field Theory, the Weyl operators  \cite{SW1,Summers87b,Weyl23} 
\begin{equation} 
	{\cal A}_f = e^{i \varphi(f)} \;, \label{Wop}
\end{equation}
where $\varphi(f)$ is the smeared scalar quantum field \cite{Haag92}, play a key role in the construction of the Von Neumann algebras \cite{Bratteli97} of localized field operators which are the main tool for  the study of the Bell-CHSH inequality. \\\\

\noindent In general, addressing the Bell-CHSH inequality by means of unitary operators requires the handling of various aspects: $i)$ provide a discussion on the physical maning of such operators, $ii)$ investigate the possible modification of the classical as well as of the quantum bound, $iii)$ provide some explicit examples to illustrate these concepts.\\\\All the above points will be elaborated in the following Sections. \\\\More precisely, in Section \eqref{Cbound} we analyse a few properties of the unitary operators and derive the classical bound. We shall see that the value $2$ will be replaced by $2\sqrt{2}$, which no longer signals violation of the Bell-CHSH inequality. An example relying on the use of the angular momentum operator will serve as an illustration  of  this change in the classical bound. In Section \eqref{Creal}  we discuss a class of unitary operators displaying the property 
\begin{equation} 
\langle \psi | \; {\cA} {\cB} \; |\psi \rangle = \langle \psi | \; {\cA}' {\cB} \; |\psi \rangle=\langle \psi | \; {\cA} {\cB}' \; |\psi \rangle=\langle \psi | \; {\cA}' {\cB}' \; |\psi \rangle = {\rm real \; value} \;. \label{realv}
\end{equation} 
It turns out that, in this case, the classical bound remains $2$. Therefore, we can safely speak of a violation of the Bell-CHSH inequality when 
\begin{equation} 
\big| \langle \psi |\; ({\cA}  + {\cA}') {\cB} + ({\cA} -{\cA}'){ \cB}' \;| \psi \rangle \big| > 2 \;. \label{real2}
\end{equation}
As an explicit realization of eqs.\eqref{realv}, \eqref{real2}, in   Section \eqref{phasespace} we shall present the Bell-CHSH inequality for phase space variables, corresponding to coordinates and momenta of a pair of particles in Quantum Mechanics. Section \eqref{Weyl}   provides the treatment of the  inequality in the case of a relativistic real scalar quantum field by means of the Weyl operators. In Section \eqref{Conclusion} 
we collect our conclusion.

\section{Properties of the unitary operators. Derivation of the classical bound}\label{Cbound}

It is worth starting this section with  a clarification. In what follows the word {\it classical} will be referred to commuting quantities, while the word {\it quantum} will be employed to denote non-commuting objects, {\it i.e.} operators. \\\\Let $\cA$ be a generic unitary operator. To capture the physical content of $\cA$ we set 
\begin{equation} 
{\cA} = {\cal M} + i {\cal N} \;, \label{capt}
\end{equation}
where $(\cal M, \cal N)$ are Hermitian operators. From ${\cA}{ \cA^{\dagger}}={\cA^{\dagger}} {\cA} =1$, one gets 
\begin{equation} 
\left[ {\cal M}, {\cal N} \right] = 0 \;, \qquad {\cal M}^2 + {\cal N}^2=1 \;.
\end{equation}
This equation tells us that a unitary operator can be thought as being made up by two commuting Hermitian operators whose eigenvalues are distributed along the unit circle. \\\\For what it concerns the classical bound, we consider four unitary commuting quantities 
$(e^{i\alpha},e^{i\alpha'})$,   $(e^{i\beta}, e^{i\beta'})$ and look at the  quantity 
\begin{equation}
{\cal Z}_{unit} = (e^{i\alpha} + e^{i\alpha'}) e^{i\beta} + (e^{i\alpha} - e^{i\alpha'}) e^{i\beta'} \;. \label{cZ}
\end{equation}
Use of the triangle inequality gives 
\begin{equation} 
\big| {\cal Z}_{unit} \big| \le  \big| e^{i\alpha} + e^{i\alpha'} \big| + \big| e^{i\alpha} - e^{i\alpha'} \big| \;. \label{tr}
\end{equation}
Therefore 
\begin{equation} 
\big| {\cal Z}_{unit} \big| \le \sqrt{2} \left( \;\sqrt{1+ \cos(\alpha-\alpha') } + \sqrt{1-\cos(\alpha-\alpha') } \;\right)  \;, \label{cos}
\end{equation}
whose maximum value is $2\sqrt{2}$, namely 
\begin{equation} 
\big| {\cal Z}_{unit} \big| \le 2\sqrt{2} \;. \label{max}
\end{equation}
Instead, in the usual case of Bell's dichotomic commuting quantities
\begin{equation}
{\cal Z}_{{dichotomic}} = (a+a')b + (a-a')b' \;,\qquad a^2={a'}^2=b^2={b'}^2 =1\;,  \label{zdic}
\end{equation}
we have the well known result 
\begin{equation}
\big| {\cal Z}_{{dichotomic}} \big| \le 2 \;. \label{maxdic}
\end{equation}
We see thus that, in the case of unitary quantities, the classical bound increases till $2\sqrt{2}$.

\subsection{Example of a correlator attaining $2\sqrt{2}$. A non-violating Bell-CHSH case}\label{nonv}
It is instructive to provide here the example of a  quantum mechanical model attaining the value of $2\sqrt{2}$  which, however,  cannot be considered a violation of the Bell-CHSH inequality. \\\\For that purpose, we rely on the orbital angular moment in two-dimensions\footnote{We set ${\hbar=1}$.}, {\it i.e.}
\begin{equation}
L = \varepsilon_{ij} x_i p_j = - i \frac{\partial }{\partial \vartheta} \;. \label{l2}
\end{equation}
For the eigenvalues and eigenvectors we have 
\begin{equation} 
L | m \rangle = m\; |m\rangle \;, \qquad |m\rangle = \frac{e^{i\; m \vartheta}}{\sqrt{2 \pi}} \;, \qquad m={\rm integer} \;, \label{eig}
\end{equation} 
and 
\begin{equation} 
\langle m\;| \; m' \rangle = \frac{1}{2\pi} \int_0^{2 \pi} d\vartheta \; e^{-i (m-m')\vartheta} = \delta_{mm'} \;. \label{norm}
\end{equation}
We consider now a bipartite system: ${\cal H}= {\cal H}_A \otimes {\cal H}_B$. The total angular momentum is 
\begin{equation} 
L_T = L_A + L_B \;, \qquad \left[L_A,L_B \right] = 0 \;. \label{Lt}
\end{equation}
As unitary operators, we choose 
\begin{equation} 
{\cA} = e^{i \alpha L_A} \;, \qquad {\cA}'= e^{i \alpha' L_A} \;, \qquad {\cB}=e^{i \beta L_B}\;, \qquad{ \cB}'= e^{i \beta L_B} \;, \label{BL}
\end{equation}
where $(\alpha, \alpha', \beta, \beta')$ are arbitrary real parameters which play a role akin to that of the four Bell's angles. Denoting by $|m_A, m_B> = |m_A> \otimes |m_B>$ the eigenvectors of $L_T$, as entangled state $|\psi>$ we shall take the state with vanishing total angular momentum given by 
\begin{equation} 
|\psi> = \frac{1}{\sqrt{2}} \left( |1,-1> - |-1,1>   \right)  \;. \label{st}
\end{equation}
An elementary calculation yields 
\begin{equation} 
\langle \psi |\; e^{i \alpha L_A} e^{i \beta L_B} \;| \psi \rangle = \cos(\alpha-\beta) \;. \label{Lc}
\end{equation}
Therefore, for the Bell-CHSH correlator, we get 
\begin{eqnarray} 
\langle C_L \rangle & = &  \langle \psi |\; (e^{i \alpha L_A}+ e^{i \alpha' L_A}) e^{i \beta L_B}+ (e^{i \alpha L_A}- e^{i \alpha' L_A}) e^{i \beta' L_B} \;| \psi \rangle \nonumber \\ 
& = &\cos(\alpha-\beta) + \cos(\alpha'-\beta) + \cos(\alpha-\beta') - \cos(\alpha'-\beta) \;. \label{CLT}
\end{eqnarray} 
Making use of the standard choice: 
\begin{equation} 
\alpha = 0\;, \qquad \alpha'= \frac{\pi}{2} \;, \qquad \beta = \frac{\pi}{4} \;, \qquad \beta'= \frac{3\pi}{4} \;, \label{stc}
\end{equation}
one ends up with 
\begin{equation} 
\langle C_L \rangle = 2\sqrt{2} \;. \label{imp}
\end{equation}
This equation might give the impression that the Bell-CHSH correlator \eqref{imp} would saturate Tsirelson's bound, attaining a maximal violation. Though,  this is an incorrect statement, the reason being that the unitary operators \eqref{BL} all commute among themselves, {\it i.e.} $[{\cA},{\cA}']=[{\cB},{\cB}']=0$, thus not fulfilling one of the basic requirements of eqs.\eqref{ABBellu}. As such, the result  \eqref{imp} does not imply  a violation of the Bell-CHSH inequality. Rather, in agreement with eq.\eqref{max}, the expression \eqref{imp} provides an example of a non-violating correlator attaining the maximum allowed classical bound: $2 \sqrt{2}$.

\section{A particular class of unitary operators}\label{Creal}

This section is devoted to a special class of unitary operators for which, instead of $2\sqrt{2}$, the classical bound returns to the usual value $2$. A relevant example of this class of unitary operators is given by the Weyl operators, a key ingredient for the  study of the violation of the Bell-CHSH inequality in relativistic Quantum Field Theory. \\\\In addition to the requirements \eqref{ABBellu}, these operators enjoy the property of displaying correlation functions which are real. More precisely, they fulfill 

\begin{equation} 
\langle \psi | \; {\cA} {\cB} \; |\psi \rangle = \langle \psi | \; {\cA}' {\cB} \; |\psi \rangle=\langle \psi | \; {\cA} {\cB}' \; |\psi \rangle=\langle \psi | \; {\cA}' {\cB}' \; |\psi \rangle = {\rm real \; value} \;. \label{realv1}
\end{equation}
Interestingly enough, the above equations have their classical counterpart, namely 
\begin{eqnarray}
Im(e^{i \alpha} \;e^{i\beta} )& = & \sin(\alpha+ \beta) = 0 \;, \qquad Im(e^{i \alpha} \; e^{i\beta'}) = \sin(\alpha + \beta') =0 \;, \nonumber \\
Im(e^{i \alpha'} \; e^{i\beta}) & = & \sin(\alpha+ \beta) = 0 \;, \qquad Im(e^{i \alpha'} \; e^{i\beta'} )= \sin(\alpha' + \beta') = 0\;, \label{imc}
\end{eqnarray}
implying 
\begin{equation}
\alpha + \beta = \pi n_1 \;, \qquad \alpha + \beta'= \pi n_ 2 \;, \qquad \alpha'+\beta= \pi m_1 \;, \qquad \alpha'+ \beta'= \pi m_2 \;, \label{ig}
\end{equation}
with $(n_1,n_2,m_1,m_2)$ integers. Thus 
\begin{equation} 
\alpha - \alpha'= \pi (n_1-m_1) = \pi \;\rm{integer} \;. 
\end{equation} 
As a consequence, the maximum value of expression \eqref{cos} becomes  now $2$, {\it i.e}
\begin{equation} 
\big| {\cal Z}_{unit} \big| \le 2 \;, \label{two}
\end{equation}
giving back the usual value.

\section{Bell-CHSH inequality violation for phase space variables}\label{phasespace}

In this section we provide a detailed analysis of how the use of unitary operators simplifies the search for violation of the Bell-CHSH inequality in the case of phase space variables. This is not a straightforward problem \cite{Cavalcanti2007}, but as we will discuss now, it can be approached with the use of  unitary operators. 

We consider the following set of Weyl operators 
\begin{align}
	A &= e^{i\alpha X_1}; \;\;\;\; A' = e^{i\alpha' P_1}; \nonumber\\
	B &= e^{i\beta X_2}; \;\;\;\; B' = e^{i\beta' P_2};
\end{align}
Where $X_i$ and $P_i$ are position and momentum operators of particle $i$. The Bell-CHSH inequality reads
\begin{align}
	\langle \psi | C |\psi \rangle = \langle \psi | \left[(A+A') B + (A-A') B')\right] |\psi \rangle 
\end{align}
It is better to work with wave functions to perform the computations. For instance
\begin{align}
	\langle \psi | AB' |\psi \rangle = \int dx_1 dx_2  e^{i\alpha x_1} \psi^{\ast}(x_1, x_2) \psi(x_1, x_2 + \beta')
\end{align}
We note that $A'$ and $B'$ act as translation operators, shifting $x_1$ and $x_2$, respectively, while $A$ and $B$ just multiplies a phase factor. Therefore, the general expressions we are interested in are of the form
\begin{align}\label{masterint}
	I = \int dx_1 dx_2  e^{i\alpha x_1} e^{i\beta x_2}\psi^{\ast}(x_1, x_2) \psi(x_1 + \alpha', x_2 + \beta')
\end{align}
The wave function $\psi(x_1, x_2 )$ must be entangled if we hope to violate the Bell-CHSH inequality.

We consider the wave function proposed by Bell \cite{bell-speak}, given by
\begin{align}
	\psi(x_1,x_2) = N \left( (x_1-x_2)^2 - 8 \sigma^2_{-} \right) e^{-\frac{(x_1-x_2)^2 }{8 \sigma^2_{-}}}e^{-\frac{(x_1+x_2)^2 }{8 \sigma^2_{+}}}
\end{align}
where $N$ is the normalization factor.

Expression \eqref{masterint} thus reads
\begin{align}\label{masterint2}
	I = N^2 \int dx_1 dx_2  e^{i\alpha x_1} e^{i\beta x_2}& \left( (x_1-x_2)^2 - 8 \sigma^2_{-} \right) \left( (x_1-x_2+ \alpha'-\beta')^2 - 8 \sigma^2_{-} \right) \nonumber\\
	& e^{-\frac{(x_1-x_2)^2 }{8 \sigma^2_{-}}}e^{-\frac{(x_1-x_2 +\alpha'-\beta')^2 }{8 \sigma^2_{-}}}e^{-\frac{(x_1+x_2)^2 }{8 \sigma^2_{+}}}e^{-\frac{(x_1+x_2 +\alpha'+\beta')^2 }{8 \sigma^2_{+}}}
\end{align}
We can recast this expression in a more convenient form by defining $x_{-} = \frac{x_1 - x_2}{\sqrt{2}}$ and $x_{+} = \frac{x_1 + x_2}{\sqrt{2}}$, so that
\begin{align}\label{masterint3}
	I = 4N^2 \int dx_{-} dx_{+}   e^{i\frac{\alpha -\beta}{\sqrt{2}} x_{-}} e^{i\frac{\alpha +\beta}{\sqrt{2}} x_{+}}&\left( x_{-}^2 - 4 \sigma^2_{-} \right)  \left(\left( x_{-} +\frac{(\alpha' -\beta')}{\sqrt{2}}\right) ^2 - 4 \sigma^2_{-} \right)\nonumber\\
	& e^{-\frac{x_{-}^2}{4 \sigma^2_{-}}}e^{-\frac{(x_{-} +\frac{(\alpha' -\beta')}{\sqrt{2}})^2 }{4 \sigma^2_{-}}}e^{-\frac{x_{+}^2}{4 \sigma^2_{+}}}e^{-\frac{(x_{+} +\frac{(\alpha' +\beta')}{\sqrt{2}})^2 }{4 \sigma^2_{+}}}
\end{align}

Defining the standard integrals, that can be evaluated by elementary methods 
\begin{align}\label{stand1}
	I_1(a,b,c)  &= \int dx e^{-ax^2} e^{-a(x+c)^2}e^{ibx} = \sqrt{\frac{\pi}{2a}}  e^{\frac{-b^2}{8a}} e^{\frac{-ac^2}{2}} e^{\frac{-ibc}{2}}  \nonumber\\
	I_2(a,b,c,d)  &= \int dx e^{-ax^2} e^{-a(x+c)^2}e^{ibx} \left(x^2 -d^2\right) \left((x+c)^2 -d^2\right)\nonumber\\
	&= \sqrt{\frac{\pi}{2a}}  e^{\frac{-b^2}{8a}} e^{\frac{-ac^2}{2}} e^{\frac{-ibc}{2}}  \left(\frac{3}{16 a^2} - \frac{3 b^2}{32 a^3} + \frac{b^4}{256  a^4}  + \left(\frac{c^2}{4}+d^2\right)\left(\frac{b^2}{8a^2} - \frac{1}{2a}\right)  +\left(\frac{c^2}{4}+d^2\right)^2 -d^2c^2\right)\nonumber\\
	&= \sqrt{\frac{\pi}{2a}}  e^{\frac{-b^2}{8a}} e^{\frac{-ac^2}{2}} e^{\frac{-ibc}{2}}  \left(\frac{3}{16 a^2} - \frac{d^2}{2a} +d^4 -  \frac{3 b^2}{32 a^3} + \frac{b^4}{256  a^4}  + \frac{b^2 d^2}{8  a^2} + \frac{b^2 c^2}{32  a^2}  - \frac{c^2}{8  a} - \frac{c^2 d^2}{2} + \frac{c^4}{16} \right)
\end{align}
we note that
\begin{align}\label{masterint4}
	I = 4N^2 I_1\left(\frac{1}{4 \sigma^2_{+}}, \frac{(\alpha +\beta)}{\sqrt{2}}, \frac{(\alpha' +\beta')}{\sqrt{2}}\right) I_2\left(\frac{1}{4 \sigma^2_{-}}, \frac{(\alpha -\beta)}{\sqrt{2}}, \frac{(\alpha' -\beta')}{\sqrt{2}}, 2 \sigma_{-}\right)
\end{align}
It follows that 
\begin{align}
	\langle \psi | AB |\psi \rangle &= 4N^2   I_1\left(\frac{1}{4 \sigma^2_{+}}, \frac{(\alpha +\beta)}{\sqrt{2}}, 0 \right) I_2\left(\frac{1}{4 \sigma^2_{-}}, \frac{(\alpha -\beta)}{\sqrt{2}}, 0, 2 \sigma_{-}\right)\nonumber\\
	\langle \psi | A'B |\psi \rangle &= 4N^2  I_1\left(\frac{1}{4 \sigma^2_{+}}, \frac{\beta}{\sqrt{2}}, \frac{\alpha'}{\sqrt{2}}\right) I_2\left(\frac{1}{4 \sigma^2_{-}}, -\frac{\beta}{\sqrt{2}}, \frac{\alpha'}{\sqrt{2}}, 2 \sigma_{-}\right)\nonumber\\
	\langle \psi | AB' |\psi \rangle &= 4N^2  I_1\left(\frac{1}{4 \sigma^2_{+}}, \frac{\alpha}{\sqrt{2}}, \frac{\beta'}{\sqrt{2}}\right) I_2\left(\frac{1}{4 \sigma^2_{-}}, -\frac{\alpha}{\sqrt{2}}, \frac{\beta'}{\sqrt{2}}, 2 \sigma_{-}\right)\nonumber\\
	\langle \psi | A'B' |\psi \rangle &= 4N^2 I_1\left(\frac{1}{4 \sigma^2_{+}}, 0, \frac{(\alpha' +\beta')}{\sqrt{2}}\right) I_2\left(\frac{1}{4 \sigma^2_{-}}, 0, \frac{(\alpha' -\beta')}{\sqrt{2}}, 2 \sigma_{-}\right)
\end{align}
It is convenient to measure dimensional quantities with respect to the natural scales $\sigma_{-} $ and $\sigma_{+}$. We thus define 
\begin{align}
	\alpha = \frac{a}{\sigma_{+}} ;\;\;\; \beta = \frac{b}{\sigma_{+}};\;\;\;  \alpha' = a'\sigma_{-} ;\;\;\;  \beta' = b'\sigma_{-}
\end{align}
such that $a, a', b, b'$ are dimensionless. Also noting that the normalization is given by 
\begin{align}
	4N^2 \left(2\pi \sigma_{-} \sigma_{+}\right) \left(11\sigma^4_{-}\right)   = 1 
\end{align}
The expressions become 
\begin{align}
	\langle \psi | AB |\psi \rangle = e^{-\frac{(a+b)^2}{4}}e^{-\frac{(a-b)^2 }{4}\left(\frac{\sigma_{-} }{\sigma_{+} }\right)^2}\left(1  + \frac{(a-b)^2}{11}  \left(\frac{\sigma_{-} }{\sigma_{+} }\right)^2 +\frac{(a-b)^4 }{44}\left(\frac{\sigma_{-} }{\sigma_{+} }\right)^4\right)  \label{e1}
\end{align}

\begin{align}
	\langle \psi | A'B |\psi \rangle  = e^{-\frac{b^2 }{4}}e^{-\frac{b^2 }{4}\left(\frac{\sigma_{-} }{\sigma_{+} }\right)^2} e^{-\frac{a'^2  }{16 }}e^{-\frac{a'^2  }{16}\left(\frac{\sigma_{-} }{\sigma_{+} }\right)^2} \left( 1+  \frac{b^2}{11} \left(\frac{\sigma_{-} }{\sigma_{+} }\right)^2 +\frac{b^4}{44}\left(\frac{\sigma_{-} }{\sigma_{+} }\right)^4+ \frac{b^2 a'^2}{88}-\frac{5a'^2}{44}+ \frac{a'^4}{352}   \right) \label{e2}
\end{align}

\begin{align}
	\langle \psi | AB' |\psi \rangle  = e^{-\frac{a^2 }{4}}e^{-\frac{a^2 }{4}\left(\frac{\sigma_{-} }{\sigma_{+} }\right)^2} e^{-\frac{b'^2  }{16 }}e^{-\frac{b'^2  }{16}\left(\frac{\sigma_{-} }{\sigma_{+} }\right)^2} \left( 1+   \frac{a^2}{11} \left(\frac{\sigma_{-} }{\sigma_{+} }\right)^2 +\frac{a^4}{44}\left(\frac{\sigma_{-} }{\sigma_{+} }\right)^4+ \frac{a^2 b'^2}{88}-\frac{5b'^2}{44}+ \frac{b'^4}{352}   \right)   \label{e3} 
\end{align}

\begin{align}
	\langle \psi | A'B' |\psi \rangle = e^{-\frac{(a'+b')^2  }{16}\left(\frac{\sigma_{-} }{\sigma_{+} }\right)^2}e^{-\frac{(a'-b')^2  }{16}}\left(1   -\frac{5(a'-b')^2 }{44} + \frac{(a'-b')^4 }{704} \right) \label{e4} 
\end{align}
Note that these are all real, so that we are working in the framework of the particular case discussed in sect. \eqref{Creal}.

In order to simplify this expression, let us suppose that Alice and Bob share a state such that the relative position of the particles is almost perfectly known ($\sigma_{-} \rightarrow 0$) and the total momentum is also almost perfectly known ($\sigma_{+} \rightarrow \infty$). This state is approximately an eigenstate of the relative position and total momentum of the particles. This is the original EPR example. We therefore consider only the leading terms in the expansion in powers of $\left(\frac{\sigma_{-} }{\sigma_{+} }\right)$. We find

\begin{align}
	\langle \psi | AB |\psi \rangle \approx e^{-\frac{(a+b)^2}{4}}
\end{align}

\begin{align}
	\langle \psi | A'B |\psi \rangle  \approx e^{-\frac{b^2 }{4}} e^{-\frac{a'^2  }{16 }}\left( 1+  \frac{b^2 a'^2}{88}-\frac{5a'^2}{44}+ \frac{a'^4}{352}   \right)
\end{align}

\begin{align}
	\langle \psi | AB' |\psi \rangle  = e^{-\frac{a^2 }{4}} e^{-\frac{b'^2  }{16 }} \left(1+   \frac{a^2 b'^2}{88}-\frac{5b'^2}{44}+ \frac{b'^4}{352}   \right)
\end{align}

\begin{align}
	\langle \psi | A'B' |\psi \rangle = e^{-\frac{(a'-b')^2  }{16}}\left(1   -\frac{5(a'-b')^2 }{44} + \frac{(a'-b')^4 }{704} \right)
\end{align}

Now if we further simplify these expressions by choosing a set of parameters such that $a = -b$ and $a' = - b'$ and also $a, a' \ll 1$, so that we can consider only the leading terms, we finally obtain for the Bell-CHSH correlation
\begin{align}
	\langle \psi | C |\psi \rangle \approx 2 - \frac{a^2}{2} + \frac{31}{88} a'^2 + {\cal O}(a^4, a'^4, a^2a'^2)
\end{align}
Therefore, it is possible to obtain a violation of the Bell-CHSH inequality if  
\begin{align}
	\frac{a'^2}{a^2} >  \frac{44}{31};    \;\;\;\; a, a'\ll 1 
\end{align}
Finally, to give a more precise idea of the size of the violation, a numerical study of  expressions \eqref{e1}-\eqref{e4} has been performed, obtaining 
\begin{equation} 
\langle \psi | C |\psi \rangle = 2.19 \label{nv}
\end{equation}

\section{Weyl operators in relativistic Quantum Field Theory and violations of the Bell-CHSH inequality}\label{Weyl}

As mentioned before, this Section is devoted to present a detailed discussion of the Weyl operators in the case of a relativistic scalar Quantum Field Theory in $(3+1)$ Minkowski spacetime. As we shall see, these operators will provide a very nice example of unitary operators fulfilling the additional requirement \eqref{realv1}. As a consequence, the classical bound is still given by the value $2$ so that a violation of the Bell-CHSH inequality occurs for values bigger than $2$. For the benefit oi the reader, we shall provide a self-contained presentation of all tools needed for the facing of the Bell-CHSH inequality in relativistic Quantum Field Theory. 

\subsection{Theoretical framework}\label{TF}

Let us start by presenting a short reminder on the canonical quantization of a free real scalar field, whose action is given by
\begin{align} 
S = \int \!\! d^4x \left[\frac{1}{2} \left(\partial_\mu \phi\right)^2  - \frac{m^2}{2} \phi^2\right].  
\end{align}
The  field $\phi$ can be expanded in terms of creation and annihilation operators as 
\begin{equation} \label{qf}
\phi(t,{\vec x}) = \int \!\! \frac{d^3 {\vec k}}{(2 \pi)^3} \frac{1}{2 \omega_k} \left( a_k e^{-ikx}  + a^{\dagger}_k e^{ikx}  \right), 
\end{equation} 
where $\omega_k=k^0=\sqrt{{\vec{k}}^2 + m^2}$. The canonical commutation relations  read
\begin{align}\label{ccr}
[a_k, a^{\dagger}_q] &= (2\pi)^3 2\omega_k \delta^3({\vec{k} - \vec{q}}), \\ \nonumber 
[a_k, a_q] &= [a^{\dagger}_k, a^{\dagger}_q] = 0, 
\end{align}
Using the above definitions, one  easily obtains the commutator between the scalar fields for arbitrary  spacetime points:
\begin{align} 
\left[ \phi(x) , \phi(y) \right] = i \Delta_{PJ} (x-y), \label{caus} 
\end{align}
where $\Delta_{PJ}(x-y) $ is the  Pauli-Jordan distribution, given by:
\begin{align}\label{PJ}
	 i \Delta_{PJ}(x-y) \!=\!\! \int \!\! \frac{d^4k}{(2\pi)^3} \varepsilon(k^0) \delta(k^2-m^2) e^{-ik(x-y)},
\end{align}
where we have made use of  $\varepsilon(x) \equiv \theta(x) - \theta(-x)$. The Pauli-Jordan distribution is Lorentz-invariant and odd under the exchange $(x-y) \rightarrow (y-x)$. Moreover,  it vanishes outside of the light cone, ensuring that measurements at space-like separated points do not interfere. As such,  the Pauli-Jordan distribution encodes the information about  locality and relativistic causality. \\\\It is a well known fact that a quantum field is an operator valued distributions \cite{Haag92} and must be suitably smeared to give well-defined operators acting on the Hilbert space. This is achieved by employing   real smooth test function with compact support $h(x) \in \mathcal {C}_{0}^{\infty}(\mathbb{R}^4)$. For the   smeared  field one has
\begin{align} \label{smqf}
\phi(h) = \int \! d^4x \; \phi(x) h(x).
\end{align}
Inserting  eq.\eqref{qf} into eq.\eqref{smqf}, we can rewrite the smeared quantum field  as $\phi(h) = a_h + a^{\dagger}_h$, where we defined the smeared version of the creation and annihilation operators by
\begin{align} 
a_h &= \int \frac{d^3 {\vec k}}{(2 \pi)^3} \frac{1}{2 \omega_k}  {\hat h}^{*}(\omega_k,{\vec k}) a_k, %\nonumber \\
% a^{\dagger}_h &= \int \frac{d^3 {\vec k}}{(2 \pi)^3} \frac{1}{2 \omega_k} {\hat h}(\omega_k,{\vec k}) a^{\dagger}_k, 
\end{align} 
where ${\hat h}(p) = \int \! d^4x \; h(x)  e^{ipx}$ is the Fourier transform of $h(x)$ and with an analogous expression for $a^\dagger_h$.
Using the smeared field, one introduces  the  Lorentz-invariant inner product in the space of the test functions with compact support, given by the smeared version of the Wightman two point function \cite{Haag92}:
\begin{align} \label{IP}
    \langle f \vert g \rangle &= \langle 0 \vert \phi(f) \phi(g) \vert 0 \rangle \nonumber \\
    &= \frac{i}{2} \Delta_{PJ}(f,g) + H(f,g) \nonumber \\
    &= \int \!\! \frac{d^3p}{(2\pi)^3} \frac{1}{2 \omega_p} f^*(p) g(p)  
\end{align}
where $ \Delta_{PJ}(f,g)$ is the smeared version of eq.\eqref{PJ} and $H(f,g)$ is the symmetric combination of the smeared fields product. Thus, we can rewrite the commutator~\eqref{caus} in its smeared version as $\left[\phi(f), \phi(g)\right] = i \Delta_{PJ}(f,g)$. In this setting, causality  is thus expressed by  $\left[\phi(f), \phi(g)\right] = 0$ when the supports of $f$ and $g$ are located in spacelike separated regions. We underline that, upon using the canonical commutation relations and the above-defined inner product~\eqref{IP}, one  finds
\begin{align}
    \left[a(f), a^\dagger(g)\right] = \langle f \vert g \rangle.
\end{align}
Following \cite{SW1,Summers87b}, let $O$ be an open set in  Minkowski spacetime and $M(O)$ be the space of test functions belonging to  $\mathcal{C}_{0}^{\infty}(\mathbb{R}^4)$ with support contained in $O$: 
\begin{align} 
	M(O) = \{ f \, \vert supp(f) \subset O \}. \label{MO}
\end{align}
One introduces  the causal complement $O'$ of the spacetime region $O$ as well as the symplectic complement $M'(O)$ of the set $M(O)$ as 
\begin{align} 
O' &= \{ y \, \vert (y-x)^2 < 0, \; \forall x \in O \}, \nonumber \\	
 M'(O) &= \{ g \, \vert  \Delta_{PJ}(g,f) =0, \; \forall f \in M(O) \}. \label{MpO}
\end{align}
With the above definitions, causality can be recasted \cite{SW1,Summers87b} by stating that $\left[ \phi(f), \phi(g) \right] = 0$ whenever    $f \in M(O)$ and $g \in M'(O)$ . Furthermore, we also recast locality as \cite{SW1,Summers87b}: $M(O') \subset M'(O)$. \\\\We introduce now the Weyl operators \cite{SW1,Summers87b,Weyl23}, a well-known class of unitary operators
\begin{align}
   { \cA}_f = e^{i \phi(f)}.
\end{align}
These operators give rise to the so-called Weyl algebra 
\begin{align}
   {\cA}_f {\cA}_g = e^{-\frac{i}{2} \Delta_{PJ}(f,g)} {\cA}_{f+g}.
\end{align}
When the supports of $f$ and $g$ are spacelike separated, the Pauli-Jordan distribution vanishes, and thus the Weyl operators behave as ${\cA}_f {\cA}_g = {\cA}_{f+g} = {\cA}_g {\cA}_f$. It should also be stressed that, unlike the quantum field $\phi(f)$,  the Weyl operators  are bounded. Upon using $\phi(f) = a_f + a^\dagger_f$ and expanding the exponential, for the vacuum expectation value of the Weyl operator one gets  
\begin{align}
    \langle 0 \vert {\cA}_f \vert 0 \rangle = e^{-\frac{1}{2} \vert\vert f \vert\vert^2} \;,\label{ffn}
\end{align}
where $\vert\vert f \vert\vert^2 = \langle f \vert f \rangle$ and $\vert 0 \rangle$ is the  Fock vacuum. \\\\We proceed now by presenting a few elements of the Tomita-Takesaki modular theory \cite{Bratteli97}, a quite important ingredient in the study of the Bell-CHSH inequality in relativistic Quantum Field Theory \cite{SW1,Summers87b,Weyl23}. To that end we define the observable algebra $\mathcal{A}(O)$ associated with the spacetime region $O$ as the von Neumann algebra obtained by taking products and linear combinations of the Weyl operators defined on $M(O)$.~\footnote{For the basic definitions adopted here, see the appendix of Ref.~\cite{Weyl23}.} It is known that, by the Reeh-Schlieder theorem~\cite{Haag92}, the vacuum state $\vert 0 \rangle$ is both cyclic and separating for the von Neumann algebra $\mathcal{A}(O)$. To introduce the Tomita-Takesaki modular theory we follow \cite{Bratteli97} and define the  anti-linear unbounded operator $S$ acting on the Von Neumann algebra $\mathcal{A}(O)$ as
\begin{align} 
	S \; a \vert 0 \rangle = a^{\dagger} \vert 0 \rangle, \qquad \forall a \in \mathcal{A}(O),  \label{TT1}
\end{align}  
from which it follows that $S^2 = 1$ and $S \vert 0 \rangle = \vert 0 \rangle$. Making use of the polar decomposition of the operator $S$ \cite{Bratteli97}, one gets:
\begin{align}
S = J  \Delta^{1/2}, \label{PD}    
\end{align} 
where $J$ is anti-unitary  and $\Delta$ is positive and self-adjoint. These modular operators satisfy the following properties~\cite{Bratteli97}: 
\begin{align} 
	J \Delta^{1/2} J &= \Delta^{-1/2}, \quad \,\,	\Delta^\dagger = \Delta, \nonumber \\
	S^{\dagger} &= J \Delta^{-1/2},  \,\,\,\,\, J^{\dagger} = J, \nonumber \\
	\Delta &= S^{\dagger} S,  \quad \,\,\,\,  J^2 = 1. \label{TTP}
\end{align}
In particular, the Tomita-Takesaki theorem \cite{Bratteli97}  states that: $J \mathcal{A}(M) J = \mathcal{A}'(M)$, that is, upon conjugation by the operator $J$, the algebra $\mathcal{A}(M)$ is mapped into its commutant $\mathcal{A'}(M)$: 
\begin{equation} 
\mathcal{A'}(M) = \{ \; a', \; [a,a']=0, \forall a \in \mathcal{A}(M) \;\}  \;. \label{commA}
\end{equation} 
The Tomita-Takesaki construction has many relevant consequences  when applied to Quantum Field Theory. As far as the Bell inequalities are concerned, it gives a way of constructing Bob's  operators from Alice's ones by making use  of the modular conjugation $J$. That is, given Alice's operator ${\cA}_f$, one can assign the operator $B_f = J {\cA}_f J$ to Bob, with the guarantee that they commute  each other since by the Tomita-Takesaki theorem the operator  ${\cB}_f = J {\cA}_f J \in \mathcal{A'}(M)$  \cite{Weyl23}.\\\\When equipped with the Lorentz invariant inner product $\langle f| g\rangle$, eq.\eqref{IP}, the set of test functions give rise to a complex Hilbert space $\mathcal{F}$ which, as outlined in \cite{Rieffel77}, enjoys many properties. It turns out that the subspaces $M$ and $iM$ are standard subspaces for $\mathcal{F}$, meaning that:  i) $M \cap i M = \{ 0 \}$; ii) $M + i M$ is dense in $\mathcal{F}$. Moreover, as proven in \cite{Rieffel77}, for standard subspaces it is possible to set a modular theory analogous to that of the Tomita-Takesaki. One introduces an operator $s$ acting on $M + iM$ as
\begin{align}
    s (f+ih) = f-ih. \;, \label{saction}
\end{align}
for $f,h \in M$. Notice that with this definition, it follows  that $s^2 = 1$. Making use of the  polar decomposition, one has:  
\begin{align}
    s = j \delta^{1/2},
\end{align}
where $j$ is an anti-unitary operator and $\delta$ is  positive and self-adjoint.  Similarly to the operators $(J, \Delta)$, the  operators $(j,\delta)$ fulfill  the following properties:
\begin{align}
    j \delta^{1/2} j &= \delta^{-1/2}, \,\,\,\,\,\,  \delta^\dagger = \delta\nonumber \\
    s^\dagger &= j \delta^{-1/2}, \,\,\, j^\dagger = j \nonumber \\
    \delta &= s^\dagger s, \,\,\,\,\,\,\,\,\, j^2=1.
\end{align}
An important result \cite{Rieffel77} concerning the operator $s$ is that it holds that a test function $f \in M$ if and only if 
\begin{equation} 
s f = f \;. \label{sff}
\end{equation}
In fact, suppose $f \in M$. On general grounds, owing to eq.\eqref{saction}, one writes 
\begin{equation}
sf = h_1 + i h_2 \;, \label{pv1}
\end{equation}
for some $(h_1,h_2)$. Since $s^2=1$ it follows that 
\begin{equation} 
f = s(h_1 + i h_2) = h_1 -i h_2  \;, \label{pv2}
\end{equation} 
so that $h_1=f$ and $h_2=0$. In much the same way, one has that $f' \in M'$ if and only if $s^{\dagger} f'= f'$. \\\\As shown in \cite{Eck},  the action of the operators $(j,\delta)$ on the von Neumann algebra $\mathcal{A}(M)$ is defined through 
\begin{align} 
J W_f J =  J e^{i {\phi}(f) } J  \equiv e^{-i {\phi}(jf) } \;, \qquad \Delta e^{i {\phi}(f) } \Delta^{-1} = e^{i {\phi}(\delta f) }\;. \label{jop}
\end{align} 
Also, it is worth noting that if $f \in M$ then $jf \in M'$. This property follows from 
\begin{equation} 
s^{\dagger} (jf) = j \delta^{-1/2} jf = \delta f = j (j\delta f) = j (sf) = jf \;. \label{jjf} 
\end{equation} 
Let us end this section by mentioning that, in the case of wedge regions of the Minkowski spacetime, the spectrum of the operator $\delta$ coincides with the positive real line, that is, $\log(\delta) = \mathbb{R}$ \cite{Bisognano75,Summers87b}, being thus an unbounded operator with a continuous spectrum.

\subsection{The Bell-CHSH inequalities}\label{BTT}

We are now ready to face the Bell-CHSH inequality in the case of a real scalar relativistic Quantum Field Theory. As done before, we first introduce the Alice operators, namely 
\begin{equation}
{\cA}_f = e^{i\phi(f)} \;, \qquad {\cA}_{f'} = e^{i\phi(f')} \;, \label{alop}
\end{equation}
where $(f,f')$ are test functions belonging to $M$. Thus, from the Tomita-Takesaki construction, Bob's operators will be defined by 
\begin{equation}
{\cB}_{jf} = e^{i a \;\phi(jf)} \;, \qquad {\cB}_{jf'} = e^{i b \;\phi(jf')} \;, \label{ab}
\end{equation}
where $(a,b)$ are arbitrary real coefficients. The unitary operators $({\cB}_{jf}, {\cB}_{jf'})$ fulfill, by construction, the property 
\begin{equation} 
\left[ {\cA}_f , {\cB}_{jf} \right] = \left[ {\cA}_{f'} , {\cB}_{jf} \right] = \left[ {\cA}_f , {\cB}_{jf'} \right]= \left[ {\cA}_{f'} , {\cB}_{jf} \right] = 0\;. \label{commz}
\end{equation}
Moreover, due to eq.\eqref{ffn}, the unitary operators operators $({\cA}_f, {\cA}_{f'}, {\cB}_{jf}, {\cB}_{jf'})$ fulfill the reality condition  \eqref{realv1}, namely: 
\begin{equation}  
\langle 0|\; {\cA}_f\; {\cB}_{jf} \; |0\rangle = e^{-\frac{1}{2}||f + ajf ||^2} = {\rm real} \;. 
\end{equation}
Similar equations hold for $( \langle 0|\; {\cA}_{f'} {\cB}_{jf} \; |0\rangle, \langle 0|\; {\cA}_f\; {\cB}_{jf'} \; |0\rangle, \langle 0|\; {\cA}_{f'}; {\cB}_{jf'} \; |0\rangle )$. \\\\Thus, for the Bell-CHSH inequality in the vacuum state, we get 
\begin{eqnarray} 
\langle 0| \; {\cal C} \; |0\rangle &=& \langle 0|\; {\cA}_f\; {\cB}_{jf}  + {\cA}_{jf}\; {\cB}_{jf}  + {\cA}_f\; {\cB}_{jf'} - {\cA}_{f'}\; {\cB}_{jf} \; |0\rangle \nonumber \\[3mm]
& =& e^{-\frac{1}{2}||f + ajf ||^2} + e^{-\frac{1}{2}||f' + ajf ||^2} +e^{-\frac{1}{2}||f + bjf' ||^2} - e^{-\frac{1}{2}||f' + bjf' ||^2} \;. \label{Bqfft}
\end{eqnarray} 
From expressions \eqref{alop}, \eqref{ab}, one sees that the Weyl operators $({\cA}_f , {\cA}_{f'}, {\cB}_{jf}, {\cB}_{jf'} )$ play the same role of the displacement operators employed in Quantum Mechanics for the construction of the coherent states. In fact, from the Hausdorff-Campbell-Baker formula, the expression for ${\cA}_f $ can be rewritten as 
\begin{equation} 
{\cA}_f  = e^{i(a_f + a^{\dagger}_f)} = e^{-\frac{||f||^2}{2}}\; e^{i a^{\dagger}_f}\; e^{i a_f} \;. \label{Hcb}
\end{equation} 
When acting on the vacuum state $|0\rangle$, the operator ${\cA}_f $ gives rise  to what is called a coherent state i Quantum Field Theory. \\\\In order to evaluate the norms $ \vert\vert f+a jf \vert\vert^2$,  $ \vert\vert f+b jf' \vert\vert^2$,  $ \vert\vert f'+ajf \vert\vert^2$, $ \vert\vert f'+bjf' \vert\vert^2$,  we follow the procedure outlined in \cite{SW1,Summers87b}. Making use of the fact that the operator $\delta$ has a continuous spectrum coinciding with the positive real line, we pick up the spectral subspace specified by $[\lambda^2-\varepsilon, \lambda^2+\varepsilon ] \subset [0.1]$. Let $\varphi$ be a normalized vector belonging to this subspace. One notices that $j\varphi$ is orthogonal to $\varphi$, {\it i.e.} $\langle \varphi |  j\varphi \rangle = 0$. In fact, from 
\begin{equation} 
\delta^{-1} (j \varphi) - j (j \delta^{-1} j) \varphi = j (\delta \varphi)  \;, \label{orth}
\end{equation}
it follows that the modular conjugation $j$ exchanges the spectral subspace $[\lambda^2-\varepsilon, \lambda^2+\varepsilon ]$ into $[1/\lambda^2-\varepsilon,1/ \lambda^2+\varepsilon ]$. Therefore, proceeding as in \cite{Weyl23}, we set 
\begin{align}
f  &= \eta (1+s) \varphi \\
f'  &= \eta' (1+s) i\varphi \;, \label{nmf}
\end{align}
where $(\eta,\eta')$ are normalization factors. Since $s^2=1$, it turns out that 
\begin{equation} 
s f = f \;, \qquad s f' = f'  \;, \label{sff}
\end{equation}
so that  both $f$ and $f'$ belong to $M$. Recalling that $\varphi$ belongs to the spectral subspace $[\lambda^2-\varepsilon, \lambda^2+\varepsilon ] $, it follows that \cite{Weyl23}, 
\begin{align}
\vert\vert f \vert\vert^2  &= \vert\vert jf \vert\vert^2 = \eta^2 (1+\lambda^2) \nonumber \\
\vert\vert f' \vert\vert^2  &= \vert\vert jf' \vert\vert^2 = \eta'^{2} (1+\lambda^2) \;. \label{sfl}
\end{align}
as well as
\begin{align} 
\langle f \vert jf \rangle &= 2 \eta^2 \lambda \nonumber \\
\langle f \vert jf' \rangle &= 0 \;, \nonumber \\
\langle f' \vert jf \rangle &= 0 \nonumber \\
\langle f' \vert jf' \rangle &= 2 \eta'^2 \lambda \;. \label{scalp} 
\end{align}
Finally, for the Bell-CHSH inequality,  one has
\vspace{1cm}
\begin{eqnarray}
\langle {\cal C} \rangle &= &  e^{-\frac{1}{2}\eta^2(  (1+\lambda)^2(1+a^2)) + 4a\lambda)  } + e^{-\frac{1}{2}( \eta'^2  (1+\lambda)^2 + a^2 \eta^2(1+\lambda^2) ) } \nonumber \\[3mm] 
&+& e^{-\frac{1}{2}( \eta'^2 b^2 (1+\lambda)^2 +  \eta^2(1+\lambda^2) ) }  - e^{-\frac{1}{2}\eta'^2(  (1+\lambda)^2(1+b^2)) + 4b\lambda)  } \label{fbchsh}
\end{eqnarray}
Despite its simplicity, this equation captures very well the violation of the Bell-CHSH inequality in Quantum Field Theory. When maximized with respect to the parameters $(\eta, \eta', \lambda,a,b)$, one gets, for example: 
\begin{equation}
\eta = 0.001\;, \qquad \eta'= 0.511 \;, \qquad a= 0.227\;, \qquad b= 0.892\;, \qquad \lambda = 0.974 \;, \label{fvls}
\end{equation} 
giving 
\begin{equation}
\langle {\cal C} \rangle = 2.189  \;, \label{BV}
\end{equation}
which represents a sensible violation of the Bell-CHSH inequality in the vacuum state of the theory \cite{SW1,Summers87b}.  \\\\It remains now to establish the quantum upper bound for the correlator \eqref{Bqfft}. In order to achieve this task we evaluate ${\cal C}^{\dagger} {\cal C}$, obtaining
\begin{eqnarray}
{\cal C}^{\dagger} {\cal C}& =& 4 - e^{-i\varphi(f+ a jf)} e^{i \varphi(f'+ bjf')} - e^{-i\varphi(f'+ b jf')} e^{i \varphi(f+ ajf)}  \nonumber \\
& +&  e^{-i\varphi(f'+ a jf)} e^{i \varphi(f+ bjf')} +- e^{-i\varphi(f+ b jf')} e^{i \varphi(f'+ajf)} \;. \label{cdc}
\end{eqnarray} 
Making use of the properties of the operator norm $||\;\;||_{op}$\footnote{The operator norm of a bounded  operator $T$ acting  on a Hilbert space ${\cal H}$ is defined as 
\begin{equation} 
||T||_{op} = sup \left( \frac{||Tx||}{||x||}\;, x \in {\cal H} \right) \;. \label{nop}
\end{equation}
}
it follows that
\begin{equation} 
||{\cal C}^{\dagger} C ||_{op}\le 8 \;. \label{eight}
\end{equation} 
Moreover, for any state $|\psi\rangle$ normalized to one, we have 
\begin{equation}
\big| \langle \psi|\; {\cal C} \; |\psi \rangle \big|^2 \le ||{\cal C}\psi||^2 = \langle \psi |\; {\cal C}^{\dagger} {\cal C}\;|\psi\rangle \le || {\cal C}^{\dagger} {\cal C} \psi||\le ||{\cal C}^{\dagger} {\cal C}||_{op} \le 8 \;. \label{fb}
\end{equation}
This equation states that the quantum bound for the Bell-CHSH correlator \eqref{Bqfft} is precisely Tsirelson's bound: 
\begin{equation} 
|\langle {\cal C} \rangle| \le 2 \sqrt{2} \;. \label{ff}
\end{equation}

\section{Conclusion}\label{Conclusion}
In this work we have pursued the discussion on the role of the normal operators in the Quantum Theory, as advocated by \cite{Hu_2017,Erhard_2020}. In particular, we have investigated the Bell-CHSH inequality in the case in which unitary operators are employed. As underlined, these operators show up in a natural way when facing the Bell-CHSH inequality in relativistic Quantum Field Theory, Sect. \eqref{Weyl}. \\\\Our findings can be summarized as follows: 
\begin{itemize}

\item for generic unitary operators fulfilling eqs.\eqref{ABBellu}, the classical bound becomes $2\sqrt{2}$. An example of this phenomenon has been provided by the quantum mechanical model of Sect. \eqref{nonv} 
\item if, in addition of eqs.\eqref{ABBellu}, the unitary operators obey the reality constraint  \eqref{realv}, the classical bound returns to the usual value of $2$. As a consequence, a violation of the Bell-CHSH inequality occurs for values bigger than $2$, as it happens in the case of the phase space variables discussed in Sect.\eqref{phasespace} and  the real relativistic scalar field discussed in Sect.\eqref{Weyl}.

\item let us conclude by underlining that the present set looks very suited for the study of the Bell-CHSH inequality for continuous variables systems, see \cite{prop,prop1,prop2,prop4}.

\end{itemize}

\section*{Acknowledgments}
	The authors would like to thank the Brazilian agencies CNPq and CAPES, for financial support.  S.P.~Sorella, I.~Roditi, and M.S.~Guimaraes are CNPq researchers under contracts 301030/2019-7, 311876/2021-8, and 310049/2020-2, respectively.

%%%%%%%%%%%%%%%%%%%%%%%%%%%%%%%%%%%%%%%%%%%%%%%%%%%%%%%%%%%%%%%%%%%%%%%%%%%%%%%%%%%%%%%%%%%%%%%%%%%%%%%%%%

\end{document}